\documentclass[aps, pra, twocolumn, amsmath, amssymb, superscriptaddress, nofootinbib]{revtex4-1}

\makeatletter
\def\p@subsection{}
\def\p@subsubsection{}
\makeatother

\usepackage[utf8]{inputenc}
\usepackage{graphicx}
\usepackage{units}
\usepackage{color}
\usepackage[pdftex, colorlinks=true, linkcolor=myblue, citecolor=myblue,
urlcolor=myblue]{hyperref}
\usepackage{times}
\usepackage{comment}
\usepackage{graphicx}
\usepackage{amsmath, calc}
\usepackage{mathrsfs}
\usepackage{amsfonts}
\usepackage{amssymb}
\usepackage{bm}
\usepackage{bbm}
\usepackage{color}
\usepackage{array}
\usepackage{units}
\usepackage{tabstackengine}
\stackMath

\definecolor{myblue}{rgb}{0,0,1}
\definecolor{myred}{rgb}{1,0,0}

\DeclareMathOperator{\Tr}{Tr}

\DeclareMathOperator{\erf}{erf}

\DeclareMathOperator{\arcsinh}{arcsinh}

\begin{document}


\title{Two-photon charging of a quantum battery with a Gaussian pulse envelope}


\author{C. A. Downing} 
\email{c.a.downing@exeter.ac.uk}
\affiliation{Department of Physics and Astronomy, University of Exeter, Exeter EX4 4QL, United Kingdom}

\author{M. S. Ukhtary}
\affiliation{Department of Physics and Astronomy, University of Exeter, Exeter EX4 4QL, United Kingdom}
\affiliation{Research Center for Quantum Physics, National Research and Innovation Agency (BRIN), South Tangerang 15314, Indonesia}


\date{\today}


\begin{abstract}
\noindent \textbf{ABSTRACT}~Quantum energy science is rapidly emerging as a domain interested in the generation, transfer and storage of energy at the quantum level. In particular, quantum batteries have the scope to exploit the wonders of quantum mechanics in order to boost their performance as compared to their electrochemical equivalents. Here we show how an exponential enhancement in stored energy can be achieved with a quantum battery thanks to a two-photon charging protocol. We consider theoretically a quantum battery modelled as a quantum harmonic oscillator, which when driven by a quadratic field (manifested by a Gaussian pulse envelope) gives rise to squeezing of the battery. This quantum squeezing ensures that the population of the battery is driven exponentially up its bosonic energy ladder. Our results demonstrate a plausible mechanism for quickly storing a vast amount of energy in a quantum object defined by continuous variables, which may be explored experimentally in a variety of quantum optical platforms.
\\
\\
\noindent \textbf{Keywords:}~ quantum battery, quantum technology, energy storage, quantum harmonic oscillator, quantum optics, squeezing.
\end{abstract}


\maketitle



\section{Introduction}
\label{Sec:Introduction}


Three years ago an influential experiment exploiting ultrafast optical spectroscopy demonstrated the impressive energy storage capacity of an organic microcavity~\cite{Quach2022}. These paradigmatic empirical results kickstarted a journey towards the creation of a quantum device enabling the temporary hoarding of energy, commonly known as a quantum battery. This pivotal experiment itself came in the wake of a pioneering theoretical proposal for a quantum battery by Alicki and Fannes back in 2013~\cite{Alicki2013}, with further and significant theory developments arising in the years immediately afterwards~\cite{Binder2015, Campaioli2017, Campaioli2023}.

Groundbreaking experiments have recently revealed how one can benefit from dark states in the charging of a battery based upon a superconducting qutrits~\cite{Hu2022, Zheng2023}, as well as probing the charging of photonic quantum batteries using single-photon interferometric networks~\cite{Qu2023}, and exploring the impact of entanglement on the charging performance of quantum batteries within a linear optical platform~\cite{Huang2023}. In parallel, the charging of quantum batteries has also been paid its due theoretical attention, including studies of the influence of various external time-dependent drives on the efficacy of the battery~\cite{Zhang2019, Chen2020, NewCrescente2020, Ukhtary2024, Gemme2024}, as well as specialist setups exploiting directional interactions~\cite{Sturges2022, Ahmadi2024}. The aim of such research is to propose and realize a charging protocol which enables the storage of large amounts of energy on a short timescale, ensuring an impressive charging power of the quantum battery device~\cite{Campaioli2023}.

Here we consider an alternative approach to charge a quantum battery, by exploiting a quadratic driving field~\cite{Savona2017, Savona2019, Campinas2023, Downing2023}. In particular, we consider a two-photon drive~\cite{Leghtas2015, Wang2019, Gaikwad2023} carrying a Gaussian pulse envelope as a prototypical example of our proposed charging protocol. The quadratic nature of the drive induces quantum squeezing in the system~\cite{Walls1983, Loudon1987}, which acts to propel the mean population of the battery exponentially up its energy ladder. In what follows we present the analytic theory underpinning our idea, including explicit expressions for the enhanced stored energy, charging time and charging power.

\begin{figure}[tb]
 \includegraphics[width=1.0\linewidth]{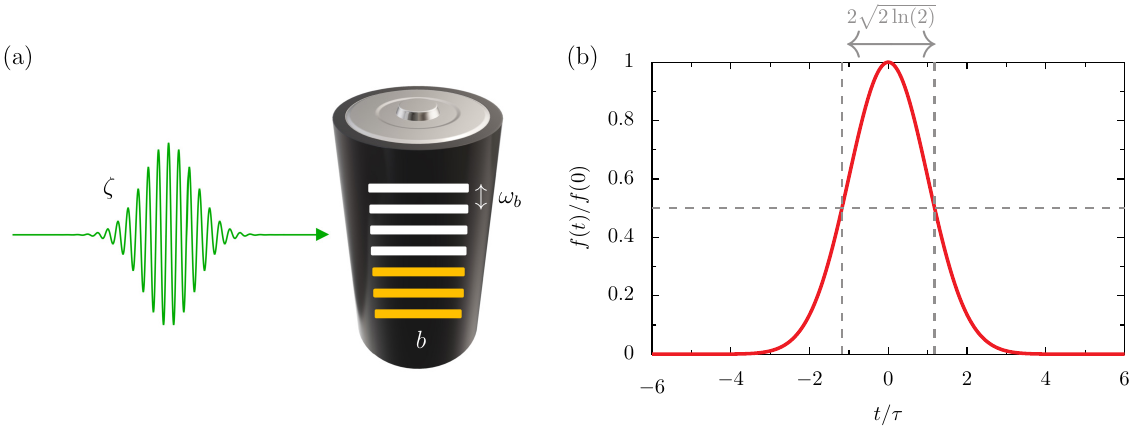}
 \caption{\textbf{Charging the quantum battery.} Panel (a): a cartoon of the quantum battery model [cf. Eq.~\eqref{eq:Haxcsdfdsfdsfsdfvcxvmy}], represented by a quantum harmonic oscillator (associated with the bosonic operator $b$) with the energy level spacing $\omega_b$ [cf. Eq.~\eqref{eq:sfdvvsssc}]. The oscillator is driven quadratically by a pulse envelope (green line) of dimensionless strength $\zeta$ [cf. Eq.~\eqref{eq:sfdcvfdddn}]. In this drawing, the battery is storing an energy $E_b = 3 \omega_b$ (three yellow bars). Panel (b): the temporal shape of the pulse envelope $f(t)$ is considered to be Gaussian [cf. Eq.~\eqref{eq:sfvraaaaa}].}
 \label{wavey}
\end{figure}


\section{Model}
\label{Sec:Model}


The total Hamiltonian operator $\hat{H}$ of the quantum battery model is comprised of just two pieces [cf. the sketch of Fig.~\ref{wavey}~(a)]
\begin{equation}
\label{eq:Haxcsdfdsfdsfsdfvcxvmy}
 \hat{H} =  \hat{H}_b + \hat{H}_{d}. 
\end{equation}
Here the bare Hamiltonian $\hat{H}_b$ describes the energy ladder of the battery (white and yellow bars in the cartoon), which is characterized by its regular energy level spacing $\omega_b$. Energy enters the battery via a pulse envelope $f(t)$, which has the dimensionality of frequency and which is carried by some cosinusoidal quadratic driving. The driving Hamiltonian $\hat{H}_d$ is also distinguished by some dimensionless strength $\zeta > 0$. Together these constituent Hamiltonians then read (we set $\hbar = 1$ throughout)
\begin{align}
  \hat{H}_b &= \omega_b \, b^\dagger b, \label{eq:sfdvvsssc} \\
   \hat{H}_d &= \zeta  \cos \left( 2 \omega_d t \right) f(t) \left( b^\dagger b^\dagger + b b \right),  \label{eq:sfdcvfdddn}
\end{align}
where the bosonic creation and annhilhiation operators $b^\dagger$ and $b$ satisfy the commutation rule $[ b, b^\dagger ] = 1$, and which act to track the scaling of excitations up and down the bosonic energy ladder. Notably, quantum harmonic oscillators are known to be excellent models for quantum optical battery systems in the real world~\cite{Qu2023}. The two-photon cosinusoidal carrier driving is of frequency $2 \omega_d$, while the pulse envelope $f(t)$ is considered to be of a Gaussian character [cf. the plot of Fig.~\ref{wavey}~(b)]
\begin{equation}
\label{eq:sfvraaaaa}
 f(t) = \frac{\mathrm{e}^{-\frac{t^2}{2\tau^2}}}{\sqrt{2 \pi \tau^2}}.
\end{equation}
Clearly, the pulse has a maximum exactly at the instant in time $t=0$, where $f(0) = 1/\sqrt{2\pi \tau^2}$, and the pulse displays a full width at half maximum of $2 \sqrt{2 \ln \left( 2 \right)} \tau \simeq 2.35 \tau$, where the temporal height and width parameter $\tau \ge 0$. The limiting case of $\lim_{\tau \to 0}  f(t) =  \delta(t)$ recovers a toy model Dirac delta function problem~\cite{Christensen1979, Downing2024}.

\begin{figure*}[tb]
 \includegraphics[width=1.0\linewidth]{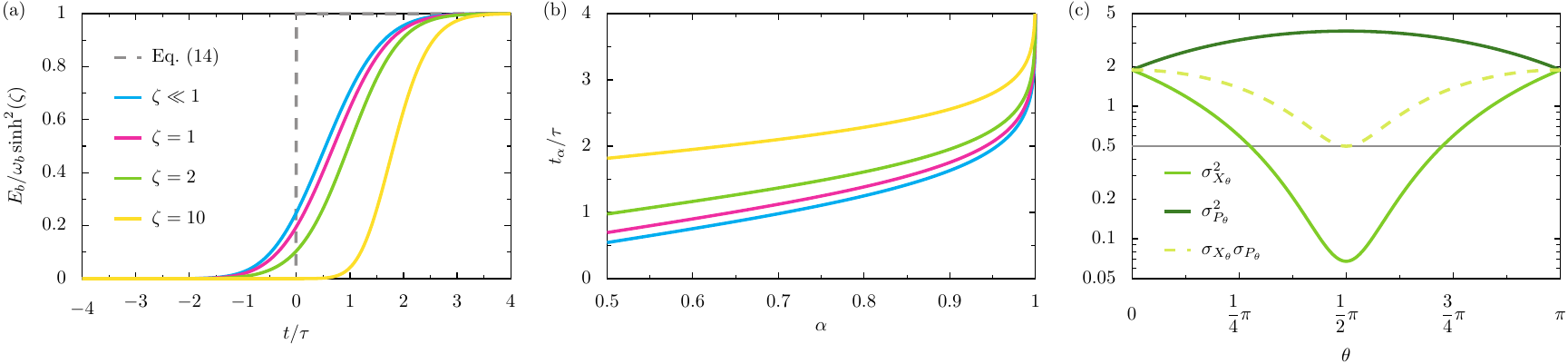}
 \caption{\textbf{Energy storage in the quantum battery.} Panel (a): The energy $E_b$ stored in the battery [cf. Eq.~\eqref{eq:csdc}] in units of the upper energetic bound $\omega_b \sinh^2 \left( \zeta \right)$, as a function of time $t$ (in units of the pulse temporal parameter $\tau$). Dashed grey line: the equivalent result in the delta pulse limit [cf. Eq.~\eqref{eq:sdfsdf}]. Panel (b): the charging time $t_{\alpha}$ (which has elapsed since the central action of the pulse at $t = 0$) at which the battery holds the fraction $\alpha$ of its energetic maximum. In panels (a) and (b), we consider several values of the dimensionless pulse strength $\zeta$ as marked in panel (a). Panel (c): a logarithmic plot of the generalized quadrature variances $\sigma_{X_\theta}^2$ (light green line) and $\sigma_{P_\theta}^2$ (dark green line) as a function of the quadrature twist angle $\theta$ [cf. Eq.~\eqref{eq:sfsddacf} and Eq.~\eqref{eq:sfqsqqssdf}]. Dashed green line: the product of the standard deviations $\sigma_{X_\theta} \sigma_{P_\theta}$. Thin grey line: a guide for the eye at $1/2$, the Robertson–Schrödinger minimum uncertainty. In panel (c) we take $\zeta = 2$ and we consider the behaviour at the snapshot in time when $t = 0$. }
 \label{energy}
\end{figure*}

The quantum state of the quantum battery may be described by a density matrix $\rho$, which evolves in time according to the Liouville–von Neumann equation $\partial_t \rho = \mathrm{i} [ \rho,  \hat{H}]$, where the Hamiltonian operator $\hat{H}$ is given by Eq.~\eqref{eq:Haxcsdfdsfdsfsdfvcxvmy}. Upon switching to a rotating frame with the aid of the time-dependent unitary transformation $U = \exp{ ( \mathrm{i} \hat{H}_b t ) }$ [cf. Eq.~\eqref{eq:sfdvvsssc}], the quantum state $\rho$ transforms into the rotated version $\tilde{\rho} = U \rho U^\dagger$, whose dynamics are instead governed by the twisted von Neumann equation
\begin{equation}
  \partial_t \tilde{\rho} = \mathrm{i} \left[ \tilde{\rho}, \hat{\mathcal{H}} \right], \label{eq:fcfefv} 
\end{equation}
where the rotated Hamiltonian operator $\hat{\mathcal{H}} = U  \hat{H}_d U^\dagger$. At driving resonance (where $\omega_d = \omega_b$) this object simplifies into
\begin{equation}
   \hat{\mathcal{H}} =  \tfrac{\zeta}{2}  f(t) \left(  b^\dagger b^\dagger + b b \right), \label{eq:sdfsfdsfcvf} 
\end{equation}
where we have also neglected the quickly rotating terms coming from the cosine function first appearing in Eq.~\eqref{eq:sfdcvfdddn}.

We are primarily interested in two figures of merit which appraise the performance of the proposed quantum battery: the dynamical energy $E_b$ stored in the battery and the instantaneous power $P_b$ of the device. These two quantities are defined by~\cite{Campaioli2023}
\begin{align}
 E_b &= \omega_b \langle b^\dagger b \rangle, \label{eq:fgdgfgd} \\
  P_b &= \partial_t E_b. \label{eq:fgasdddgfgd}
\end{align}
The stored energy $E_b$ is counted using the mean population $\langle b^\dagger b \rangle$ of the battery. For example, in the cartoon of Fig.~\ref{wavey}~(a) the three yellow bars represent the storing of an energy $E_b = 3 \omega_b$ in the quantum battery. Meanwhile, the instantaneous power $P_b$ is a popular measure of the charging prowess of quantum batteries~\cite{Salamon2020, Kamin2020, Rosa2022}. A third quantity, the so-called ergotropy $\mathcal{E}$, describes how much work can be extracted from the quantum battery~\cite{Alicki2013, Allahverdyan2004}. The thermodynamic efficiency of the proposed model of Eq.~\eqref{eq:Haxcsdfdsfdsfsdfvcxvmy} is perfect ($E_b = \mathcal{E}$) since the quantum harmonic oscillator is only coherently pumped (it starts in the vacuum state at $t \to - \infty$ and then it evolves forwards in time according to the quadratic Hamiltonian of Eq.~\eqref{eq:Haxcsdfdsfdsfsdfvcxvmy}, such that it remains in a coherent state at all times, with zero passive state energy). This thermodynamic point is discussed further in Ref.~\cite{Farina2019} and in the Supplementary Material.


\section{Results}
\label{Sec:Results}


The mean population of the system $\langle b^\dagger b \rangle$ can be readily computed using the trace propery $\Tr{(O \tilde{\rho})} = \langle O \rangle$ for an arbitrary operator $O$, where the quantum state $\tilde{\rho}$ is defined using Eq.~\eqref{eq:fcfefv} and Eq.~\eqref{eq:sdfsfdsfcvf}. Carrying out this tracing process leads to the following equation of motion for the three second moments of the quantum battery
\begin{equation}
\label{eq:sdfsfd}
\mathrm{i} \partial_t \Psi = \mathcal{M} \Psi + \mathcal{P},
\end{equation}
where the mean values of the operators are housed within $\Psi$, the driving term is $\mathcal{P}$ and the dynamical matrix is $\mathcal{M}$. These three objects are together defined by
\begin{equation}
\label{eq:sdfssdsdfsdffd}
\Psi = \begin{pmatrix}
\langle b^\dagger b \rangle \\
\langle b b \rangle \\
\langle b^\dagger b^\dagger \rangle \\
\end{pmatrix}
\quad\quad
\mathcal{P} = \begin{pmatrix}
0 \\
\zeta f(t) \\
-\zeta f(t)\\
\end{pmatrix},
\end{equation}
\begin{equation}
\label{eq:sdfssffsddsdfsdffd}
\setstackgap{L}{1.1\baselineskip}
\fixTABwidth{T}
\mathcal{M} = \parenMatrixstack{
    0 & -\zeta f(t) & \zeta f(t)   \\
    2\zeta f(t) & 0 & 0  \\
    -2\zeta f(t) & 0 & 0
}.
\end{equation}
The exact solution of Eq.~\eqref{eq:sdfsfd} can be found using the variation of parameters method. By applying the boundary condition of all of the second moments being zero when the driving pulse is vanishing (that is, when $t \to - \infty$), the exact solution in terms of a special function read $\langle b^\dagger b \rangle = \sinh^2 ( \zeta [ 1 + \erf{(y)}  ] /2 )$, $\langle b b \rangle = -\mathrm{i} \sinh ( \zeta [ 1 + \erf{(y)}  ]  )/2$ and its complex conjugate $\langle b^\dagger b^\dagger \rangle = \mathrm{i} \sinh ( \zeta [ 1 + \erf{(y)}  ]  )/2$, with the scaled time variable $y = t/\sqrt{2}\tau$. Here we have employed the so-called error function $\erf{ \left( z \right)}$, as defined by its integral representation
\begin{equation}
\label{eq:sdfdsfdsf}
\erf{ \left( z \right)} =  \frac{2}{\sqrt{\pi}} \int_0^z \mathrm{e}^{- x^2} \mathrm{d}x.
\end{equation}
Hence, using the definition of Eq.~\eqref{eq:fgdgfgd}, the total energy $E_b$ stored in the quantum battery over time is described by
\begin{equation}
\label{eq:csdc}
 E_b = \omega_b \sinh^2 \left( \tfrac{\zeta}{2} \left[ 1 + \erf{ \left( \tfrac{t}{\sqrt{2} \tau} \right) } \right] \right). 
\end{equation}
This short expression features a remarkable exponential enhancement in energy storage (as compared to batteries driven by linear fields~\cite{Ukhtary2024}), rendering the proposed charging protocol as potentially interesting for future experimental study. The energetic formula of Eq.~\eqref{eq:csdc}, scaled by the ($t \to \infty$) asymptotic energetic maximum $\omega_b \sinh^2 ( \zeta )$, is plotted as a function of time in Fig.~\ref{energy}~(a). The coloured curves highlight the smooth step-like profile of the stored energy $E_b$ for several values of the dimensionless drive strength $\zeta$. In the extremely sharp pulse ($\tau \to 0$) limit, where the envelope profile $f(t)$ of Eq.~\eqref{eq:sfvraaaaa} behaves like a Dirac delta function, Eq.~\eqref{eq:csdc} reduces to the even simpler form
\begin{equation}
\label{eq:sdfsdf}
\lim_{\tau \to 0}  E_b = \omega_b \sinh^2 \left( \zeta \right) \Theta \left( t \right),
\end{equation}
where $\Theta \left( t \right)$ is the Heaviside step function. This straightforward limit is plotted by the dashed grey line in Fig.~\ref{energy}~(a).

Clearly from Fig.~\ref{energy}~(a), the maximum possible stored energy is only achieved asymptotically (when $t \to \infty$). This fact stems from the unboundedness (from above) of the energy levels defining the quantum battery, since we employ the infinite energy ladder of a quantum harmonic oscillator in our model [cf. Eq.~\eqref{eq:sfdvvsssc}]. In practice, this infinity is not problematic and a physical realization will have a large but finite number of energy levels. The charging time $t_{\alpha}$ at which the battery holds the fraction $0 \le \alpha \le 1$ of its energetic maximum is given by the inversion of Eq.~\eqref{eq:csdc} as
\begin{equation}
\label{eq:sdfsfddsfgs}
t_\alpha = \sqrt{2} \tau \erf^{-1}{ \left(  \frac{2}{\zeta} \arcsinh \left[ \sqrt{\alpha \sinh^2 \left( \zeta \right)} \right] - 1 \right)}.
\end{equation}
This charging time $t_\alpha$ (measuring the elapsed time since the central action of the pulse at $t = 0$) is plotted in Fig.~\ref{energy}~(b) for several values of $\zeta$ [those corresponding to panel (a)]. In particular, the limiting behaviours of Eq.~\eqref{eq:sdfsfddsfgs} at very small and very large $\zeta$ are encapsulated by the formulae
\begin{align}
\lim_{\zeta \ll 1}  t_\alpha &= \sqrt{2} \tau \erf^{-1}{ \left( 2\sqrt{\alpha} - 1\right) },  \label{eq:sdfsdfsdfsfd} \\
\lim_{\zeta \gg 1}  t_\alpha &= \tau \sqrt{ \ln \left( \frac{2}{ \pi \ln^2 \left( \alpha \right) }  \frac{\zeta^2}{ \ln \left( \frac{2 \zeta^2}{ \pi \ln^2 \left( \alpha \right) } \right) }  \right)}. \label{eq:brvr}
\end{align}
When $\zeta \ll 1$ and Eq.~\eqref{eq:sdfsdfsdfsfd} holds [cyan line in panel (b)] one notices that $t_\alpha \sim \tau$ for a large range of energetic fractions $\alpha$ (at least, away from $\alpha = 1$). When $\zeta \gg 1$, such that Eq.~\eqref{eq:brvr} is a good approximation for large enough $\alpha$ [yellow line in panel (b)], the logarithmic character of this charging time $t_\alpha$ is most apparent in the curves of Fig.~\ref{energy}~(b).

\begin{figure*}[tb]
 \includegraphics[width=1.0\linewidth]{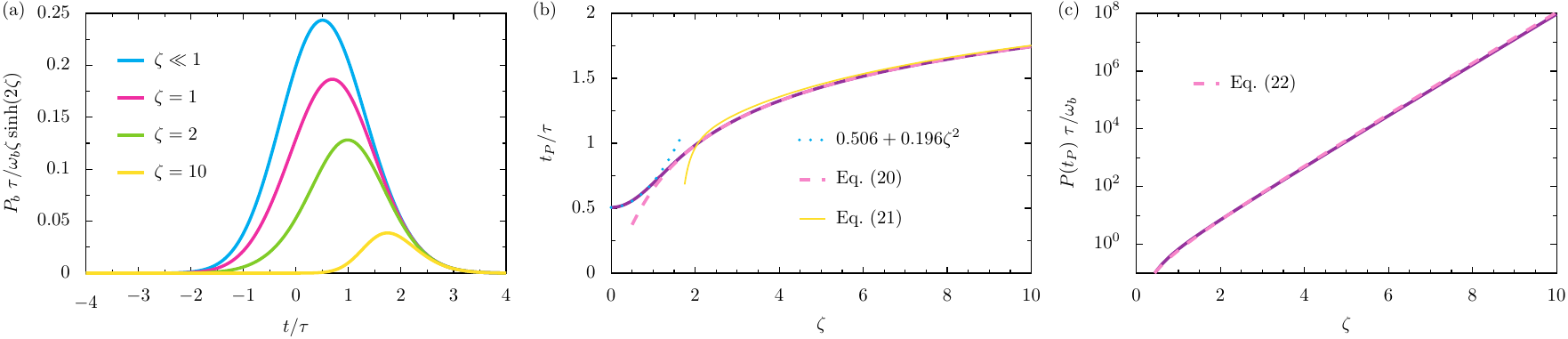}
 \caption{\textbf{Power in the quantum battery.} Panel (a): The instantaneous power $P_b$ [cf. Eq.~\eqref{eq:asdasd}], in units of $\omega_b \zeta \sinh \left( 2 \zeta \right) / \tau$, as a function of time $t$ (in units of the pulse temporal parameter $\tau$). We consider several values of the dimensionless pulse strength $\zeta$ as marked by the plot legend. Panel (b): the time $t_{P}$ (measured from the central action of the pulse at $t = 0$) at which the power $P_b$ is at its maximum as a function of $\zeta$. Dotted cyan line: a quadratic approximation in the regime $\zeta \ll 1$ [cf. Eq.~\eqref{eq:asddasd}]. Dashed pink line: the asymptotic expression when $\zeta \gg 1$ [cf. Eq.~\eqref{eq:asdadaddasd}]. Thin yellow line: a logarithmic approximation for $\zeta \gg 1$ [cf. Eq.~\eqref{eq:asddsfdsfvasd}]. Panel (c): the maximum power $P(t_{P})$, in units of $\omega_b / \tau$, as a function of $\zeta$. Dashed pink line: the approximate expression of Eq.~\eqref{eq:sdfscdscdfsdf}. }
 \label{power}
\end{figure*}

The instantaneous power $P_b$ of the quantum battery follows directly from the definition of Eq.~\eqref{eq:fgasdddgfgd} and the result of Eq.~\eqref{eq:csdc} as
\begin{equation}
\label{eq:asdasd}
P_b =  \frac{\omega_b}{\tau} \frac{\zeta}{\sqrt{2\pi}} \sinh \left( \zeta \left[ 1 + \erf{ \left( \tfrac{t}{\sqrt{2} \tau} \right) } \right] \right) \mathrm{e}^{- \frac{t^2}{2\tau^2}}.
\end{equation}
We plot the power $P_b$, in units of $\omega_b \Omega \sinh \left( 2 \zeta \right) / \tau$, as a function of time in Fig.~\ref{power}~(a) for the same spread of dimensionless pulse strengths $\zeta$ as in Fig.~\ref{energy}~(a, b). Naturally, the exponential enhancement in the stored energy is likewise found in the power $P_b$ (please note the $\sinh \left( 2 \zeta \right)$ scaling in the unit used in the plot). Larger values of $\zeta$ act to delay the moment of maximum power as shown in Fig.~\ref{power}~(a). The turning points of Eq.~\eqref{eq:asdasd} suggest that the maximum power occurs at the instant in time $t = t_P$, such that $\max_t{\{P_b\}} = P_b(t_P)$. This key time $t_P$ is plotted as function of the dimensionless pulse strength $\zeta$ with the purple line in Fig.~\ref{power}~(b). An asymptotic analysis of the transcendental turning point equation yields the limiting behaviours at small and large $\zeta$ as follows
\begin{align}
\lim_{\zeta \ll 1} t_P &\simeq 0.506 \tau, \label{eq:asddasd} \\
\lim_{\zeta \gg 1} t_P &= \tau \sqrt{W \left( \tfrac{2 \zeta^2}{\pi} \right)}. \label{eq:asdadaddasd}
\end{align}
When $\zeta \ll 1$ the peak power occurs at the specific delay time $t_P \simeq 0.506 \tau$, and this lower temporal bound grows quadratically in $\zeta$ from then on, as is sketched with the dotted cyan line in Fig.~\ref{power}~(b). Conversely, when $\zeta \gg 1$ the time of peak power nicely follows Eq.~\eqref{eq:asdadaddasd}, which is given in terms of the Lambert $W$-function (sometimes called the product logarithm function). This special function is defined as the inverse function of the equation $f(W) = W \exp{(W)}$. The slowly growing formula of Eq.~\eqref{eq:asdadaddasd} is plotted with the dashed pink line in Fig.~\ref{power}~(b), and it can be reasonably approximated using de Bruijn's expansion in terms of simple logarithms (thin yellow line) as
\begin{equation}
\label{eq:asddsfdsfvasd}
\lim_{\zeta \gg 1} t_P \simeq \tau \sqrt{ \ln{ \left( u \right) } - \ln{ \left( \ln{ u } \right) } + \tfrac{\ln{ \left( \ln{ u } \right) }}{\ln{ \left( u \right) }} },
\end{equation}
with the dimensionless variable $u = 2 \zeta^2 / \pi$. This underlines the rapid charging response of the proposed quantum battery. The maximum power $P_b(t_P)$ of the quantum battery is similarly plotted as the purple line in Fig.~\ref{power}~(c), highlighting the extremely large charging powers which can be generated thanks to the exponential enhancement induced in this model. The result can be nicely approximated by the expression
\begin{equation}
\label{eq:sdfscdscdfsdf}
P \left( t_{P} \right) \simeq \frac{\omega_b}{\tau} \frac{\mathrm{e}^{2 \left( \zeta - 1/3 \right) }}{4} \sqrt{W \left( \tfrac{2 \zeta^2}{\pi} \right)},
\end{equation}
which is marked by the dashed pink line in panel~(c). The formula of Eq.~\eqref{eq:sdfscdscdfsdf} matches well the orders of magnitude increase in maximum instantaneous power with growing dimensionless drive strength $\zeta$.

It is important to charge the device as fast as is practicable, which necessitates a maximization of the average power $\bar{P}$ (stored energy per unit time) of the quantum battery~\cite{Ferraro2018, Rosa2024}. This average charging power $\bar{P}$ can be estimated here by considering the amount of energy $E_b(t)$ deposited during some time interval $t_- \le t \le t_+$ [cf. Eq.~\eqref{eq:csdc}]. It is convenient to consider the two times $t_{\pm} = \pm \sqrt{ 2 \ln(2) } \tau \simeq \pm 1.18 \tau$, which correspond to the temporal boundaries which define the full width at half maximum of the driving pulse envelope [cf. Fig.~\ref{wavey}~(b)]. With this choice of time interval, the average charging power $\bar{P} \simeq \omega_b \sinh (\zeta) \sinh (\zeta \erf{ [ \sqrt{ \ln(2) } ] } ) / ( 2 \sqrt{ 2 \ln(2) } \tau )$ features a highly desirable exponential enhancement, as is by now familiar due to the proposed charging protocol. This charging boost is most easily seen when the drive strength is strong enough ($\zeta \gg 1$), such that the average power reduces to the explicitly exponential expression $\bar{P} \simeq \omega_b \exp{ [  \zeta ( 1 +  \erf{ [ \sqrt{ \ln(2) } ] }  ) ]} / ( 8 \sqrt{ 2 \ln(2) } \tau ) $.

The quantum squeezing induced by the quadratic nature of the pulse envelope [cf. Eq.~\eqref{eq:sfdcvfdddn}] can be better understood by considering the generalized quadrature variance $\sigma_{X_\theta}^2 = \langle \hat{X}_\theta^2 \rangle -  \langle \hat{X}_\theta \rangle^2$ and its partner quantity $\sigma_{P_\theta}^2 = \langle \hat{P}_\theta^2 \rangle -  \langle \hat{P}_\theta \rangle^2$~\cite{Walls1983, Loudon1987}. The two operators of generalized position and momentum are defined by $\hat{X}_\theta = ( \exp{ ( \mathrm{i} \theta / 2)} b^\dagger + \exp{ ( - \mathrm{i} \theta / 2)} b ) / \sqrt{2}$ and $\hat{P}_\theta = \mathrm{i} ( \exp{ ( \mathrm{i} \theta / 2)} b^\dagger - \exp{ ( - \mathrm{i} \theta / 2)} b ) / \sqrt{2}$ respectively, where $\theta$ is an angle allowing for a twist of the quadratures. Together, these operators satisfy the canonical commutation relation $[ \hat{X}_\theta, \hat{P}_\theta ] = \mathrm{i}$. The exact solution of Eq.~\eqref{eq:sdfsfd} then leads (back in the original frame) to the twin expressions
\begin{align}
\sigma_{X_\theta}^2 &= \tfrac{1}{2} +  \sinh^2 \left( \tfrac{\zeta \xi}{2} \right) - \tfrac{1}{2} \sin \left( 2\omega_b t + \theta \right)  \sinh \left( \zeta \xi \right), \label{eq:sfsddacf} \\
\sigma_{P_\theta}^2 &= \tfrac{1}{2} +  \sinh^2 \left( \tfrac{\zeta \xi}{2} \right) + \tfrac{1}{2} \sin \left( 2\omega_b t + \theta \right)  \sinh \left( \zeta \xi \right), \label{eq:sfqsqqssdf}
\end{align}
where the variable $\xi = 1 + \erf{ ( t / \sqrt{2} \tau) }$ also carries a time dependency. We plot the variances $\sigma_{X_\theta}^2$ (light green line) and $\sigma_{P_\theta}^2$ (dark green line) in Fig.~\ref{energy}~(c) for the driving strength $\zeta = 2$, and at the instant coinciding with the peak of the pulse ($t = 0$). The plot showcases the $\hat{X}_\theta$--squeezed nature of the system, since $\sigma_{X_\theta}^2$ dips below the Robertson–Schrödinger minimum uncertainty of $1/2$ (thin grey line). Meanwhile the dashed green line in Fig.~\ref{energy}~(c) confirms that the product of standard deviations always obeys the celebrated rule $\sigma_{X_\theta} \sigma_{P_\theta} \ge 1/2$. These important (and phase-dependent) quantities can be readily probed using homodyne detection experiments~\cite{Shaked2018, Porto2018}. This brief quadrature analysis serves to confirm the role of the dimensionless pulse strength $\zeta$ as a kind of squeezing parameter for the proposed quantum battery model [cf. Eq.~\eqref{eq:sfsddacf} and Eq.~\eqref{eq:sfqsqqssdf}].

In order to support our idea of exploiting quantum squeezing in order to boost the performance of quantum batteries, we have built a minimal working theory by considering a closed quantum system [cf. Eq.~\eqref{eq:sfdvvsssc} and Eq.~\eqref{eq:sfdcvfdddn}]. Of course, the effect of the external environment can also be important for the functioning of quantum batteries~\cite{Farina2019, Barra2019, Carrega2020, Seah2021, Smirne2023}, although -- quite remarkably -- the expected performance degradation of quantum batteries as open quantum systems can be stabilized in certain circumstances~\cite{Gherardini2020, Munro2020, Mitchison2021, Morrone2023}. For our purposes here, we have checked that the exponential enhancement in stored energy which we have reported is indeed robust against the inclusion of dissipation, as we show in the Supplementary Material. We have also listed some auxiliary analytical results for other pulse functions $f(t)$ (including for hyperbolic secant, Pöschl–Teller, algebraic and Lorentzian shapes) in the Supplementary Material to complement the Gaussian choice of Eq.~\eqref{eq:sfvraaaaa}.


\section{Conclusion}
\label{Sec:Conclusion}


In conclusion, we have suggested a charging protocol for a quantum battery which gives rise to an exponential enhancement in both stored energy and charging power. The main idea is to utilize a quadratic driving field, here primarily modelled with a Gaussian pulse envelope, in order to induce quantum squeezing into the system. This squeezing in turn boosts the mean population of the bosonic battery in an energetically (and ergotropically) favourable manner. We hope that our theoretical proposal can stimulate further experimental work in the area of quantum energy storage~\cite{Hu2022, Zheng2023, Qu2023, Huang2023}, with an overall view towards advancing the nascent field of quantum energy science~\cite{Campaioli2023, Metzler2023}. 
\\


\section*{Acknowledgements}
\label{Sec:Acknowledgements}

\noindent
CAD is supported by the Royal Society via a University Research Fellowship (URF\slash R1\slash 201158) and by Royal Society Enhanced Research Expenses which support MSU. CAD also gratefully acknowledges an Exeter-FAPESP SPRINT grant with the Universidade Federal de São Carlos, and thanks R.~Bachelard, A.~Cidrim and C.~J.~Villas-Boas for fruitful discussions. 
\\


\noindent \textbf{DECLARATION OF COMPETING INTERESTS}\\
The authors have no conflicts to disclose.
\\


\noindent \textbf{DATA AVAILABILITY}\\
No data were generated or analyzed in the presented research.
\\


\noindent
\textbf{ORCID.}\\
C. A. Downing: \href{https://orcid.org/0000-0002-0058-9746}{0000-0002-0058-9746}.
\\
M. S. Ukhtary: \href{https://orcid.org/0000-0001-5197-7354}{0000-0001-5197-7354}.
\\



\end{document}